\documentclass[conference]{IEEEtran}
\usepackage[T1]{fontenc}
\usepackage[utf8]{inputenc}
\usepackage{newtxtext}
\usepackage{microtype}
\usepackage[top=0.7in, bottom=1in, left=0.625in, right=0.625in]{geometry}
\IEEEoverridecommandlockouts
\usepackage{cite}
\usepackage{amsmath,amssymb,amsfonts}
\usepackage{algorithmic}
\usepackage{graphicx}
\usepackage{textcomp}
\usepackage{xcolor}
\usepackage[cmintegrals]{newtxmath}
\usepackage{bm}
\usepackage{enumitem}
\usepackage{dblfloatfix}
\usepackage{fixltx2e}

\usepackage{siunitx}

\usepackage{float}
\usepackage{subfig}
\usepackage{adjustbox}
\usepackage{booktabs}
\usepackage{tabularx}
\usepackage{makecell}

\usepackage{caption}
\usepackage{array}
\usepackage{multirow}

\newcolumntype{K}{>{\raggedright\arraybackslash}m{2.2 cm}}
\newcolumntype{E}{>{\centering\arraybackslash}m{2.3 cm}}
\newcolumntype{Y}{>{\centering\arraybackslash}m{4.2 cm}}
\newcolumntype{C}{>{\centering\arraybackslash}m{3.1 cm}}
\newcolumntype{d}{>{\centering\arraybackslash}m{1.35 cm}}
\newcolumntype{a}{>{\centering\arraybackslash}m{1.2 cm}}
\newcolumntype{e}{>{\centering\arraybackslash}m{1.35 cm}}

\newcolumntype{D}{>{\centering\arraybackslash}m{3 cm}}
\newcolumntype{z}{>{\centering\arraybackslash}m{0.7 cm}}
\newcolumntype{L}{>{\centering\arraybackslash}m{1 cm}}

\newcolumntype{x}{>{\centering\arraybackslash}m{0.75 cm}}
\newcolumntype{P}{>{\centering\arraybackslash}m{1.5 cm}}

\def\BibTeX{{\rm B\kern-.05em{\sc i\kern-.025em b}\kern-.08em
    T\kern-.1667em\lower.7ex\hbox{E}\kern-.125emX}}

\usepackage{fancyhdr}
\fancypagestyle{firststyle}{\fancyhf{}
		\fancyhead[L]{D. Shakya, M. Ying, T. S. Rappaport, P. Ma, Y. Wang, I. Al-Wazani, Y. Wu, and H. Poddar ``Upper Mid-Band Channel Measurements and Characterization at 6.75 GHz FR1(C) and 16.95 GHz FR3 in an Indoor Factory Scenario",\textit{ IEEE International Conference on Communications (ICC), }Montreal, Canada, Jun. 2025, pp. 1--6.}
	}
\usepackage{times}

\begin{document}

\title{Upper Mid-Band Channel Measurements and Characterization at 6.75 GHz FR1(C) and 16.95 GHz FR3 in an Indoor Factory Scenario}

	\author{\IEEEauthorblockN{Mingjun Ying$^{\dagger1}$, Dipankar Shakya$^{\dagger2}$, Theodore S. Rappaport$^{\dagger3}$, \\Peijie Ma$^\dagger$, Yanbo Wang$^\dagger$, Idris Al-Wazani$^\dagger$, Yanze Wu$^\dagger$, and Hitesh Poddar*}
		\IEEEauthorblockA{\textit{$^\dagger$NYU WIRELESS, Tandon School of Engineering, New York University, USA}\\
			\IEEEauthorblockA{\textit{*Sharp Laboratories of America (SLA), Vancouver, Washington, USA}}
			\{$^1$yingmingjun, $^2$dshakya, $^3$tsr\}@nyu.edu}
		\thanks{This research is supported by the New York University (NYU) WIRELESS Industrial Affiliates Program.}
	}

\maketitle

\thispagestyle{firststyle}

\begin{abstract}
This paper presents detailed radio propagation measurements for an indoor factory (InF) environment at 6.75 GHz and 16.95 GHz using a 1 GHz bandwidth channel sounder. Conducted at the NYU MakerSpace in the NYU Tandon School of Engineering campus in Brooklyn, NY, USA, our measurement campaign characterizes the radio propagation in a representative small factory with diverse machinery and open workspaces across 12 locations, comprising five line-of-sight (LOS) and seven non-line-of-sight (NLOS) scenarios. Analysis using the close-in (CI) free space path loss (FSPL) model with a 1 m reference distance reveals path loss exponents (PLE) below 2 in LOS at 6.75 GHz and 16.95 GHz, while in NLOS, PLE is similar to free-space propagation (e.g., PLE = 2). The RMS delay spread (DS) decreases at higher frequencies with a clear frequency dependence. Also, measurements show a wider RMS angular spread (AS) in NLOS compared to LOS at both frequency bands, with a decreasing trend as frequency increases. These observations in a dense-scatterer factory environment demonstrate frequency-dependent behavior that differs from existing industry-standard 3GPP models. Our findings provide crucial insights into complex propagation mechanisms in factory environments, essential for designing robust air interface and industrial wireless networks at the upper mid-band FR3 spectrum.

\end{abstract}


\section{Introduction} \label{sec:intro}

The increasing demand for higher data rates, lower latency, and enhanced reliability in wireless communication systems has driven the exploration of new frequency bands, such as FR1(C) (4-8 GHz) and FR3 (7-24 GHz). These upper mid-band frequencies offer a promising middle ground between FR1 (410 to 7125 MHz) and FR2 (24.25 GHz to 52.6 GHz) bands, providing wider bandwidth than the Sub-6 GHz spectrum while avoiding some of the propagation challenges associated with millimeter-wave (mmWave) frequencies. Regulatory agencies, such as ITU and FCC, have identified these bands between 6-24 GHz as key candidates for future 5G and 6G deployments in urban and industrial environments \cite{NTIA2024, Kang2024OJCOM}.

While significant research has focused on the higher mmWave and sub-terahertz (sub-THz) bands \cite{Rappaport2019ia, Ju21jsac, Ju2019icc, rappaport2012rws, Nie2013pimrc}, there remains a limited understanding of propagation characteristics in industrial environments. Unlike urban outdoor or indoor office spaces, factories include dense arrangements of metallic surfaces, heavy machinery, and dynamic obstacles, creating complex multipath propagation with unique reflection and scattering characteristics. Understanding these propagation mechanisms is essential for developing robust wireless systems for industrial automation, monitoring, and control applications \cite{Cheng18a, Ju21jsac}.


Previous studies demonstrated unique propagation behavior in industrial environments. Rappaport et al. \cite{rappaport1987phdthesis, rappaport1989characterization} conducted pioneering measurements at 1.3 GHz in factory settings that contributed to Wi-Fi standardization, reporting a path loss exponent (PLE) of 2.15, using the close-in (CI) free space path loss (FSPL) model with a 2.3 m (10$\lambda$) reference distance, highlighting the need for dedicated channel models for factory environments.
Vijayan et al. \cite{vijayan20235g} compared measured path loss in a medical device facility against the 3GPP InF model for sparse clutter and high-height transmitter. They found that measured path loss values at 3.6 GHz were consistently 2-4 dB lower than what the 3GPP model predicted, indicating some agreement between the measurement results and 3GPP models in \cite{3GPPTR38901}.


Al-Samman et al. \cite{al2021wideband} conducted measurements at 107-109 GHz across three industrial environments, reporting PLEs between 1.6 and 2.0 using the CI FSPL model with a 1 m reference distance. Their findings revealed that most signal energy is concentrated in early delay bins, demonstrating favorable propagation characteristics for industrial environments that support the potential deployment of future 6G networks in manufacturing settings. \textcolor{black}{Ju et al.~\cite{Ju2023twc} conducted measurements in four different factories with multiple antenna heights at 142 GHz, and provided comprehensive channel statistics. Their study revealed LOS PLE ranging from 1.6 to 1.9 in LOS using the CI FSPL model with a 1 m reference distance and a substantial increase in delay spread in the NLOS scenario.}

Wang et al. \cite{wang2024multi} studied industrial environments at 28, 38, 132, and 220 GHz, showing frequency-dependent propagation with increasing LOS PLEs with frequency, from 1.41 at 28 GHz to 1.70 at 132 GHz using the CI FSPL model with a 1 m reference distance. Lyczkowski et al. \cite{lyczkowski2021power} examined industrial automated guided vehicle lanes from 2.1 GHz to 5.1 GHz, reporting increasing NLOS PLEs from 1.35 at 2.1 GHz to 3.08 at 5.1 GHz using the CI FSPL 1 m reference distance model.


This paper presents a detailed empirical study of channel measurements at 6.75 GHz and 16.95 GHz within the NYU MakerSpace factory environment, and offers channel statistics such as path loss, delay spread, angular spread, and impact of polarization. 
The remainder of this paper is organized as follows: Section \ref{sec:system_and_scenario} describes the channel sounder, the indoor factory environment, and the measurement procedures. Section \ref{sec:chan_char} presents a comprehensive analysis of channel characteristics at 6.75 GHz and 16.95 GHz, including path loss modeling for co-polarized and cross-polarized antenna configurations, directional and omnidirectional path loss models, a comparison of delay and angular spreads across different environments and frequencies, and a comparison with the standardized values from 3GPP TR 38.901 \cite{3GPPTR38901}. Finally, Section \ref{conclusion} summarizes the key findings. 


\section{Channel Measurement in InF with a 1 GHz Bandwidth Channel Sounder}
\label{sec:system_and_scenario}

	\begin{figure}[t!]
		\centering
\includegraphics[width=0.45\textwidth]{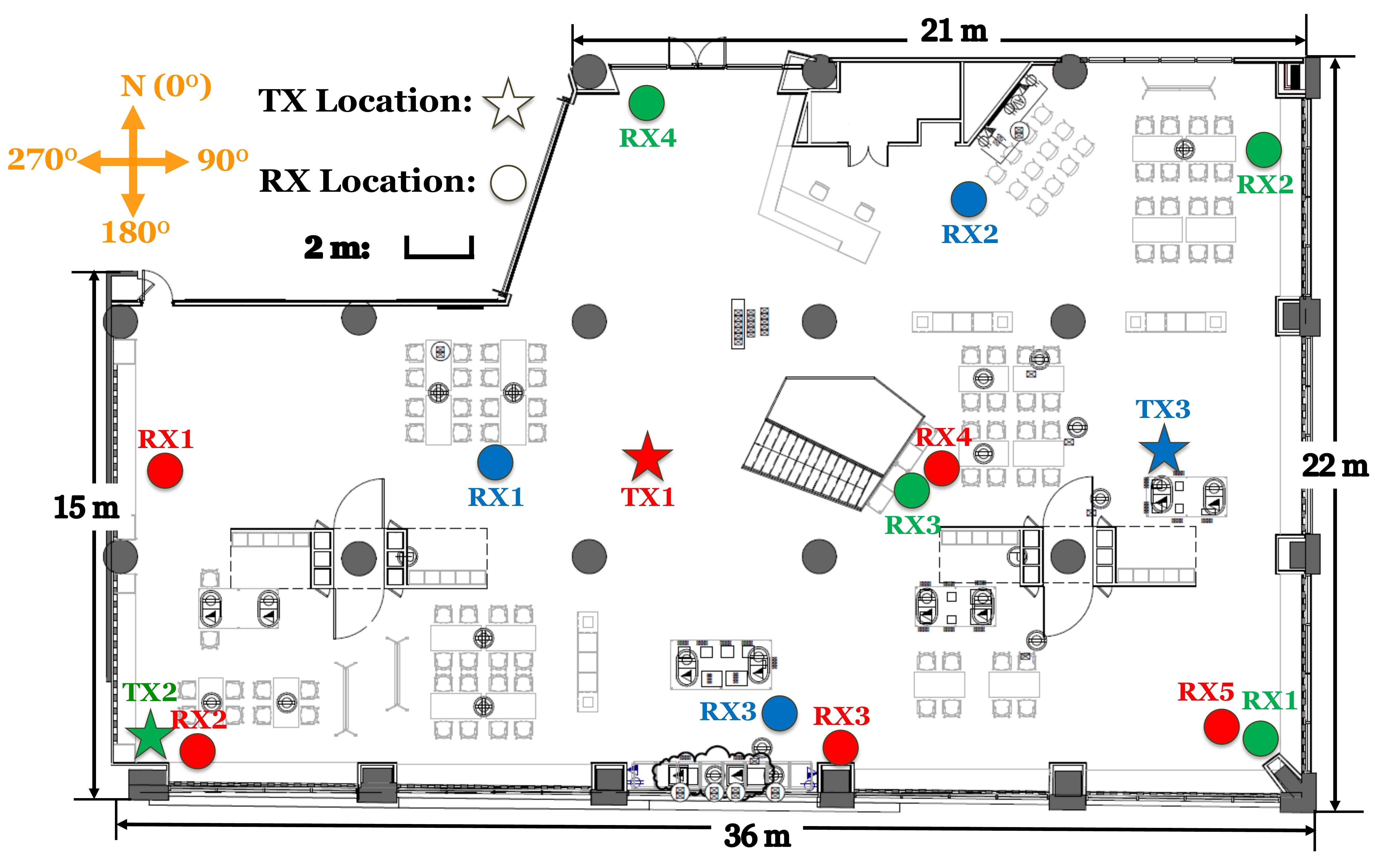}
		\caption{Floor plan of NYU MakerSpace factory environment showing three TX locations ($\bigstar$) and their associated RX positions (\scalebox{1.5}{$\bullet$}), with matching colors indicating TX-RX pairs.\newline}
		\label{fig:IndoorMap}
		\vspace{-25 pt}
	\end{figure}

\subsection{Channel Sounding System}

A sliding correlation-based wideband channel sounder, detailed in \cite{Shakya2021tcas, Mac2017jsac}, was used for radio propagation measurements. A 500 Mcps PN sequence at baseband is upconverted to a 1 GHz wideband RF signal at 6.75 GHz for FR1(C), and further upconverted to 16.95 GHz for FR3 using a heterodyne architecture. The RX employed a slightly slower 499.9375 Mcps PN sequence, resulting in a time-dilated PDP due to the sliding correlation~\cite{Ju2023twc,rappaport2012rws,Ju2019icc}.

The unique feature of this sounder, as highlighted in \cite{Shakya2024ojcoms}, is its dual-band co-located design, which allows seamless switching between 6.75 GHz and 16.95 GHz by merely changing the IF signal cable to the appropriate RF module. \textcolor{black}{During measurements, our system records the PDP for each direction over meticulous angular sweeps of directional horn antennas at both the TX and RX, where 20 instantaneous PDPs are averaged for each spatial pointing direction~\cite{sun2014icc, rappaport2012rws}.  Measurement data is stored along with the angular position of the rotating antenna gimbals and relevant operating parameters~\cite{Mac2017jsac}.} Detailed specifications of the sounder system are provided in \cite{Shakya2024gc, Shakya2024gc2, Shakya2024ojcoms}.


\subsection{Calibration of the Channel Sounder System}
\label{sxn:Cal}
The calibration procedure ensures accurate power, delay, and direction capture of multipath components (MPC). The complete calibration comprises three steps:

\subsubsection{Linearity and Power Calibration}
Transmit power was measured using a Keysight N1913A power meter with N8487A sensor, to keep within FCC's 35 dBm limit. Linaerity and power calibration used a 4 m separation with 1.5 m TX/RX heights, ensuring a single LOS path \cite{Xing2018vtc}. Power calibration was performed twice daily to maintain measurement accuracy.

\subsubsection{Time Calibration}
To ensure accurate multipath delay measurement, Rubidium (Rb) clocks at TX and RX are initially synchronized with a physical cable, then maintained via Precision Time Protocol (PTP) over a WiFi link during measurements \cite{Shakya2023gc}, eliminating post-processing or ray-tracing corrections typically needed to correct timing drift in captured data. After linearity calibration, the LOS peak in PDPs is circular-shifted to match the expected free-space propagation delay at 4 m. Re-calibration after measurement at each location compensates for timing drift during extended sessions.

\subsubsection{Spatial Calibration}
At each TX-RX location, antennas are mounted at 2.4 m (TX) and 1.5 m (RX) heights, using geographical North as the 0$^\circ$ reference for spatial coordination. 

\subsection{Indoor Factory Environment}


The NYU MakerSpace Factory, approx. \SI{36}{\meter} $\times$ \SI{22}{\meter} $\times$ \SI{5}{\meter} (W$\times$L$\times$H), is an active manufacturing space equipped with a variety of machinery, including several 3D printers, CNC machines, laser cutters, PCB fabrication and soldering stations, metal and woodwork stations, and a few robotic arms. The exterior walls of the factory on the north, south, and east sides were covered by large glass panels framed with metal, creating reflective surfaces, while the west wall consisted of thick drywall. 

During channel measurements, 3 TX locations and 12 RX locations were used, with the TX placed at a height of 2.4 m, and the RX at 1.5 m, \textcolor{black}{representing environment with dense clutter and a high base station (InF-DH)}. Concrete pillars (represented as gray-filled circles in Fig. \ref{fig:IndoorMap}) were loacated throughout the space, impacting the propagation. Alongside the pillars, various machines and material surfaces contributed to the complex scattering environment~\cite{Ju2019icc}.


The TX-RX location pairs for the 6.75 GHz FR1(C) and 16.95 GHz FR3 measurements are listed in Table \ref{tab:LSPs}, including five LOS and seven NLOS TX-RX pairs, with separations ranging from 9 m to 38 m. Our measurement system has a link margin of 155.6 dB at 6.75 GHz and 159.2 dB at 16.95 GHz\cite{Shakya2024ojcoms, Ted2025icc, Shakya2025wcnc, Shakya2025icc, Ying2025tcom}, but much less TX power was required in this campaign \cite{Ying2025tcom}. There were no outages at any InF measurement locations. Attenuation was capped to maintain linear RX response.

\vspace{-5 pt}


\subsection{Measurement Procedure}
The measurements at the TX-RX locations follow a systematic approach~\cite{Ying2025tcom, Shakya2024ojcoms}. For example,
\begin{itemize}
    \item Rapid RX azimuth scans identify significant TX angles of departure (AODs) \cite{Shakya2024ojcoms}.
    \item Detailed RX azimuthal sweeps in half-power beamwidth (HPBW) steps conducted for each AOD identified by the rapid scans following Table \ref{tab:sweeps}.
    \item Measurements in both co-polarized and cross-polarized antenna configurations~\cite{sun2014icc, rappaport2012rws, maccartney2015ia}.
    \item Calibration before and after measurements at each location for temporal and spatial accuracy.
\end{itemize}

\begin{table}[!t]
    \centering
    \caption{TX/RX elevation configurations for 360$^{\circ}$ azimuthal sweeps at fixed TX angles \cite{Shakya2024ojcoms}}
    \renewcommand{\arraystretch}{1.2}\resizebox{0.45\textwidth}{!}{
\begin{tabular}{|>{\centering\arraybackslash}m{1cm}|>{\centering\arraybackslash}m{2.4cm}|p{4.8cm}|}
        \hline
		\multicolumn{1}{|c|}{\textbf{Sweep \#}} & 
		\multicolumn{1}{c|}{\textbf{TX elevation}} & 
		\multicolumn{1}{c|}{\textbf{RX elevation}} \\
        \hline
	   \multirow{3}{*}[5pt]{1}& \multirow{3}{=}{}          & RX at \textit{boresight$^{\mathsection}$} elevation, then swept 360$^{\circ}$ in azimuth plane in HPBW$^\dagger$ steps. \\ \cline{1-1} \cline{3-3}
        \multirow{3}{*}[5pt]{2} &\multirow{3}{=}[3pt]{\centering TX at \textit{boresight$^{\mathsection}$} elevation. } & RX \textit{tilted down by one HPBW$^\dagger$}, then swept 360$^{\circ}$ in azimuth plane in HPBW steps. \\ \cline{1-1} \cline{3-3}
        \multirow{3}{*}[5pt]{3} & & RX \textit{tilted up by one HPBW$^\dagger$}, then swept 360$^{\circ}$ in azimuth plane in HPBW steps. \\
        \hline
        \multirow{3}{*}[5pt]{4} & \multirow{3}{=}[5pt]{\centering TX \textit{tilted down by one HPBW$^\dagger$.}}  & RX at \textit{boresight$^{\mathsection}$} elevation, then swept 360$^{\circ}$ in azimuth plane in HPBW steps. \\
        \hline
        \multirow{3}{*}[5pt]{5} & \multirow{3}{=}[5pt]{\centering TX \textit{tilted down by one HPBW$^\dagger$.}}  & RX \textit{tilted down by one HPBW$^\dagger$}, then swept 360$^{\circ}$ in azimuth plane in HPBW steps. \\
        \hline
    \end{tabular}}
    \par\vspace{1ex}
    \raggedright
	\scriptsize{$^\dagger$HPBW: 30$^{\circ}$ @ 6.75 GHz; 15$^{\circ}$ @ 16.95 GHz. $^{\mathsection}$Boresight: antenna orientation where TX and RX are pointed directly at each other, typically requiring slight uptilt/downtilt when at different heights \cite{sun2014icc, rappaport2012rws, maccartney2015ia, Nie2013pimrc}.}
    \label{tab:sweeps}
    \vspace{-15 pt}
\end{table}
\renewcommand{\arraystretch}{1.0}

\section{6.75 and 16.95 GHz InF Propagation}
\label{sec:chan_char}

\subsection{Path Loss}
The measured path loss obtained from the directional and omnidirectional PDPs is fit to the CI FSPL model with a 1 m free space reference distance, $d_0$, to evaluate the PLE ($n$). 
{\small
\begin{equation}
	\label{eq:CI}
	\begin{aligned}
		PL^{CI}(f_c,d_{\text{3D}})\;\text{[dB]} &= \text{FSPL}(f_c, 1 m)+ 10n\log_{10}\left( \dfrac{d_{3D}}{d_{0}} \right)+\chi_{\sigma}, \\
		\text{FSPL}(f_c,1 m) &= 32.4 + 20\log_{10}\left(\dfrac{f_c}{1\;\text{GHz}}\right),
	\end{aligned}
\end{equation}
}
where FSPL$(f_c, 1 \;\text{m})$ is obtained for carrier frequency $f_c$ GHz at 1 m, $n$ is the PLE, and $\chi_{\sigma}$ is large-scale shadow fading (zero-mean Gaussian r.v. with s.d. $\sigma^{CI}$ in dB) \cite{Rappaport2015tc}.

\renewcommand{\arraystretch}{1.6}
\begin{table*}[htbp]
    \centering
    \caption{Directional and Omnidirectional CI FSPL model with a 1 m reference distance (FI model for 3GPP) parameters for 6.75 GHz FR1(C) and 16.95 GHz FR3 InF campaigns, with comparison to 3GPP~\cite{3GPPTR38901} and 142 GHz factory campaign in \cite{Ju2023twc}}
        \vspace{-5pt}
    \scriptsize
    \resizebox{0.94\textwidth}{!}{
    \begin{tabular}{|@{\hspace{2 pt}}c@{\hspace{2 pt}}|@{\hspace{2 pt}}c@{\hspace{2 pt}}|@{\hspace{2 pt}}c@{\hspace{2 pt}}|c|p{0.32 cm}|p{0.32 cm}|c|p{0.4 cm}|p{0.32 cm}|c|p{0.32 cm}|p{0.4 cm}|p{0.32 cm}|p{0.32 cm}|p{0.4 cm}|p{0.32 cm}|p{0.32 cm}|p{0.4 cm}|}
        \cline{1-18}
        \multirow{4}{2.1cm}{\centering \textbf{Campaign}} & 
        \multirow{4}{1.1cm}{\centering \textbf{Distance (m)}} & 
        \multirow{4}{1.2cm}{\centering \textbf{Antenna HPBW (TX/RX)}} & 
        \multicolumn{9}{c|}{\textbf{Directional path loss}} & 
        \multicolumn{6}{c|}{\textbf{Omni path loss}} \\
        \cline{4-18}    
        & & & \multicolumn{3}{c|}{LOS} & \multicolumn{3}{c|}{NLOS Best} & \multicolumn{3}{c|}{NLOS} & \multicolumn{3}{c|}{LOS} & \multicolumn{3}{c|}{NLOS} \\
        \cline{4-18}    
        & & & \textbf{n} & \textbf{$\sigma$ (dB)} & \textbf{XPD (dB)} & \textbf{n} & \textbf{$\sigma$ (dB)} & \textbf{XPD (dB)} & \textbf{n} & \textbf{$\sigma$ (dB)} & \textbf{XPD (dB)} & \textbf{n} & \textbf{$\sigma$ (dB)} & \textbf{XPD (dB)} & \textbf{n} & \textbf{$\sigma$ (dB)} & \textbf{XPD (dB)} \\
        \cline{1-18}
        \centering \textbf{6.75 GHz (This work)} & 9-38 & (30$^{\circ}$/30$^{\circ}$) & 
        1.64 & 1.44 & 19.8 & 2.22 & 3.88 &                 
                    18.3 & 2.92 & 7.69 & 13.7 & 
        1.39 & 1.86 & 17.4 & 1.78 & 2.46 & 17.3 \\
        \cline{1-18}
        \centering 6.75 GHz (3GPP)~\cite{3GPPTR38901} & 10-600 & -- & 
        -- & -- & -- & -- & -- & -- & -- & -- & -- & 
        \multicolumn{3}{c|}{\parbox{2cm}{\centering $\alpha$:31.84, $\beta$:2.15, $\sigma$:4.3}} & 
        \multicolumn{3}{c|}{\hspace{0cm}\parbox{2.6cm}{\centering $\alpha_{SH}$:32.4, $\alpha_{DH}$:33.63 \\ $\beta_{SH}$:2.3, $\beta_{DH}$:2.19 \\  $\sigma_{SH}$:5.9, $\sigma_{DH}$:4}}\\
        \cline{1-18}
        \centering \textbf{16.95 GHz (This work)} & 9-38 & (15$^{\circ}$/15$^{\circ}$) & 
        2.09 & 6.2 & 25.1 & 2.48 & 4.2 & 22 & 3.81 & 10.42 & 19.1 & 
        1.75 & 3.1 & 24 & 2.11 & 3.29 & 22.8 \\
        \cline{1-18}
        \centering 16.95 GHz (3GPP)~\cite{3GPPTR38901} & 10-600 & -- & 
        -- & -- & -- & -- & -- & -- & -- & -- & -- & 
        \multicolumn{3}{c|}{\parbox{2cm}{ \centering $\alpha$:31.84, $\beta$:2.15,  $\sigma$:4.3}} & 
        \multicolumn{3}{c|}{\hspace{0cm}\parbox{2.6cm}{\centering $\alpha_{SH}$:32.4, $\alpha_{DH}$:33.63 \\ $\beta_{SH}$:2.3 , $\beta_{DH}$:2.19 \\ $\sigma_{SH}$:5.9, $\sigma_{DH}$:4}} \\
        \cline{1-18}
        \centering 142 GHz \cite{Ju2023twc} & 5-87 & (8$^{\circ}$/8$^{\circ}$) & 
        2.1 & 1.8 & 27.1 & 3.12 & 7.1 & 22.2 & 4.92 & 11.3 & 13.3 & 
        1.86 & 2.2 & 27.3 & 2.74 & 5.1 & 19 \\
        \cline{1-18}
    \end{tabular}
    }
    \label{tab:PLEs}
\par\vspace{1ex}
\parbox{\textwidth}{\raggedright \footnotesize{\textbf{Notation:} SH = Sparse clutter and High base station height (Tx or Rx elevated above the clutter), DH = Dense clutter and High base station height.}}
\vspace{-23pt}
\end{table*}
\renewcommand{\arraystretch}{1}
	
	\begin{figure}[!t]
		\centering%
		\subfloat[]{%
		\includegraphics[width=96mm]{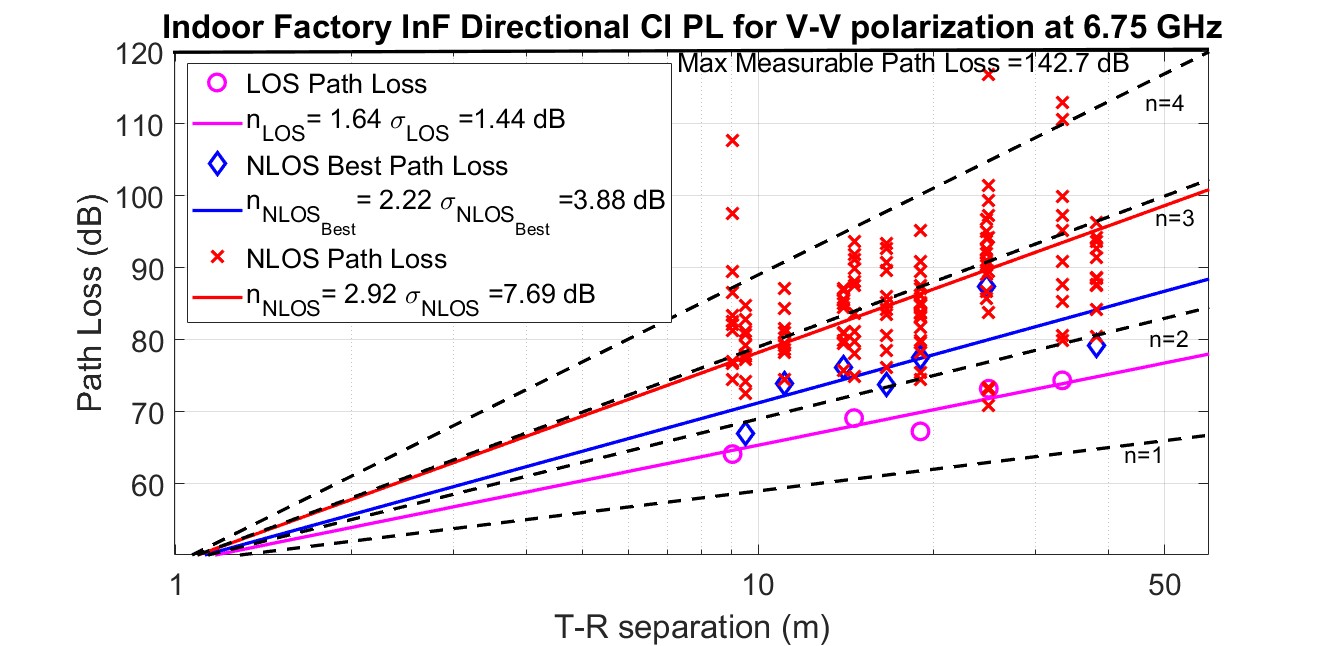}
        }\\
		\vspace{-7 pt}
		\subfloat[]{%
		\includegraphics[width=96mm]{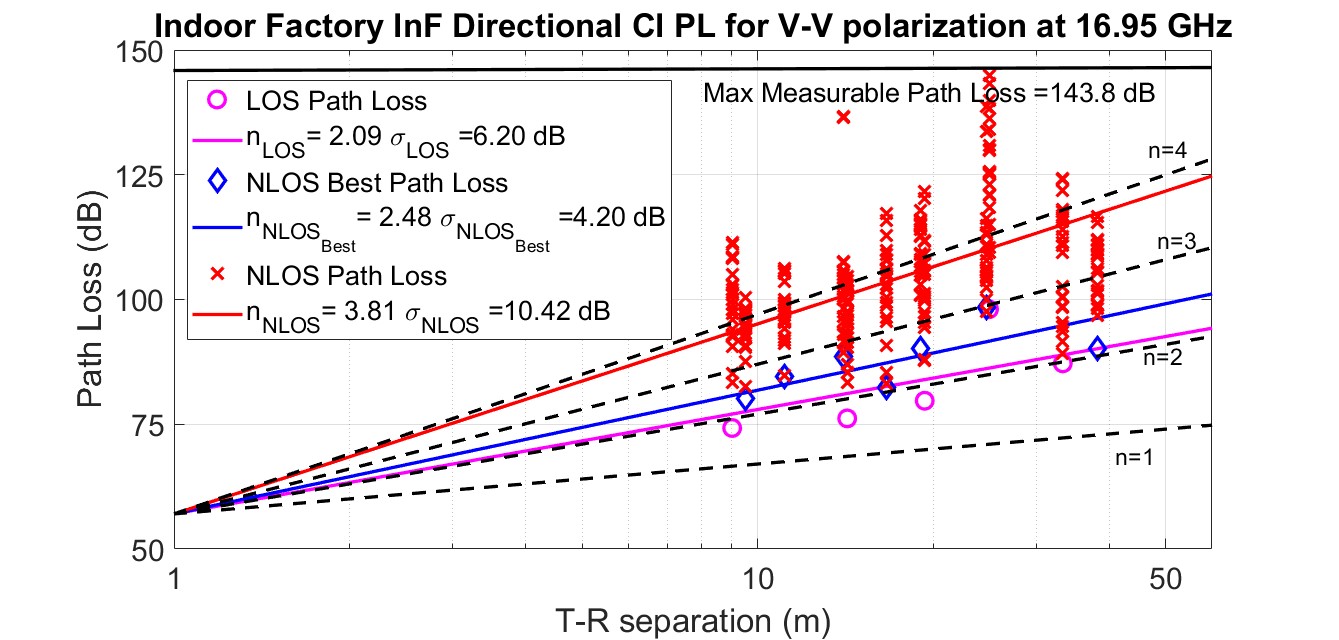}
        }
		\caption{InF directional CI FSPL models with a 1 m reference distance and scatter plot for V-V polarization at: (a) 6.75 GHz FR1(C); (b) 16.95 GHz FR3. [T-R separation: 9--38 m]}
		\label{fig:DirPL}
		\vspace{-10 pt}
	\end{figure}

    \begin{figure}
		\centering%
		\subfloat[]{%
			\centering
			\includegraphics[width=96mm]{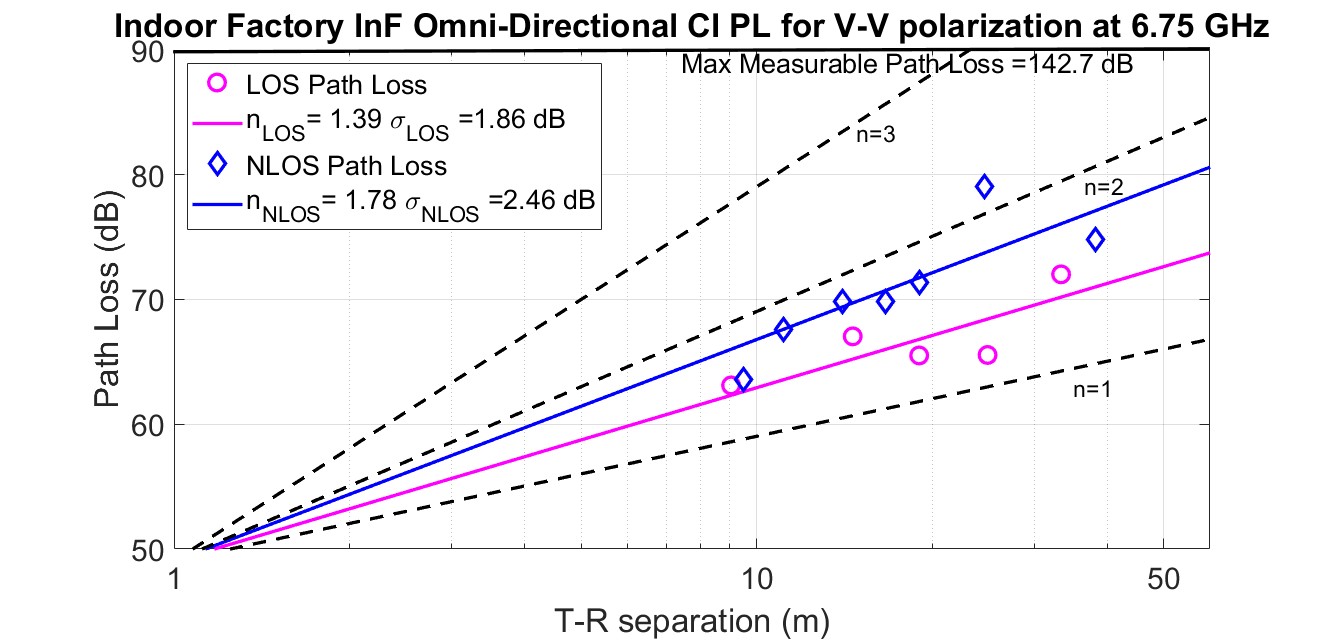}
		}\\
		\vspace{-7 pt}
		\subfloat[]{%
			\centering
			\includegraphics[width=96mm]{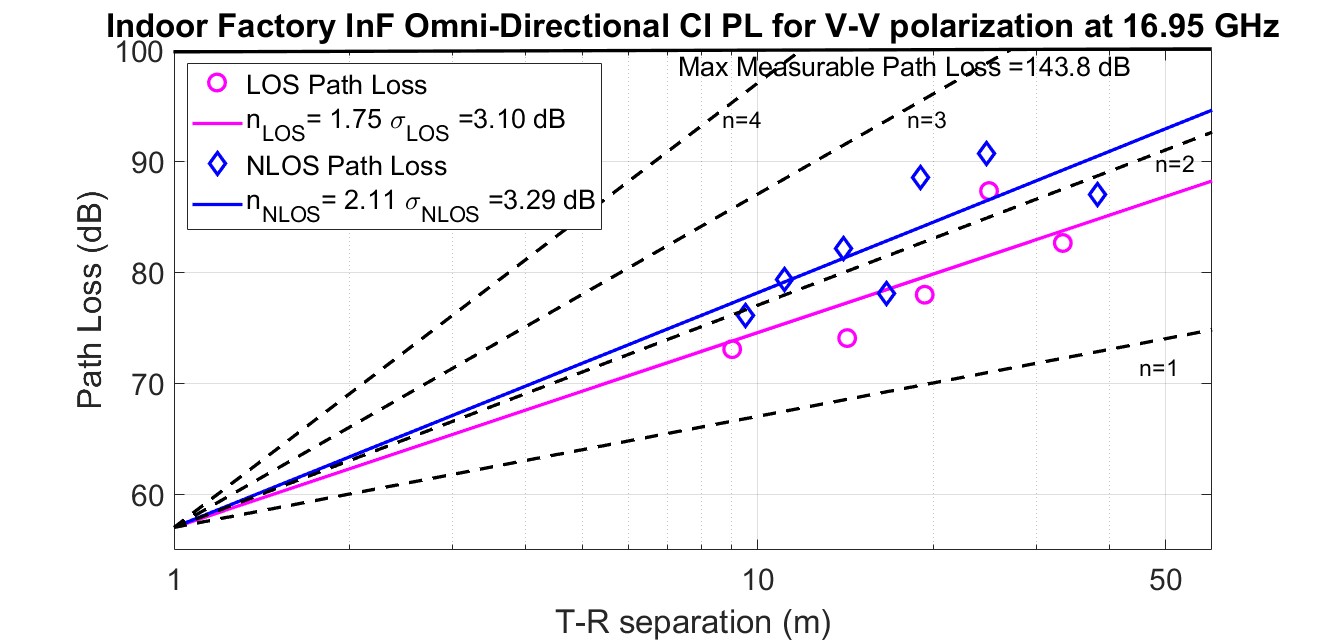}
		}%
		\\
		\caption{InF omnidirectional CI FSPL models with a 1 m reference distance and scatter plot for V-V polarization at: (a) 6.75 GHz FR1(C); (b) 16.95 GHz FR3. [T-R separation: 9--38 m]}
		\label{fig:OmniPL}
    \end{figure}

Fig. \ref{fig:DirPL} shows the directional CI FSPL models with a 1 m reference distance and scatter plots for co-polarized antennas in an InF environment at 6.75 GHz and 16.95 GHz, covering T-R separations from 9 to 38 m. The PLE is evaluated for LOS, NLOS$_{\text{Best}}$, and NLOS scenarios. NLOS$_{\text{Best}}$ PL is calculated for the NLOS RX pointing in the direction of the highest received power~\cite{maccartney2015ia}. The LOS PLE at 6.75 GHz is 1.64, with NLOS$_{\text{Best}}$ and NLOS PLEs are 2.22 and 2.92. At 16.95 GHz, PLEs of 2.09, 2.48, and 3.81 are seen for LOS, NLOS$_{\text{Best}}$, and NLOS.


After synthesizing the omni PDPs using the method described in \cite{sun2015gc, sun2014icc}, in Fig. \ref{fig:OmniPL}, we observe a lower PLE at 6.75 GHz, measuring 1.39 in LOS and 1.78 in NLOS, compared to 16.95 GHz, where the PLE values are 1.75 in LOS and 2.11 in NLOS. The low PLE at 6.75 GHz may be attributed to the structural characteristics of the MakerSpace, which includes plasterboard or wooden partitions with open sections at the top or bottom, facilitating easier penetration of lower-frequency waves.
The relatively low PLE in NLOS at 6.75 GHz suggests a quasi-line-of-sight (QLOS) scenario, as defined in \cite{otim2019impact}, where partial obstructions allow for some signal diffraction around obstacles, yielding a PLE less than 2, unlike fully obstructed NLOS conditions. Additionally, the absence of permanent walls contributes to the reception of scattered power from multiple directions at NLOS locations. The frequency comparison in Table \ref{tab:PLEs} shows lower PLE at FR3 frequencies compared to sub-THz (e.g., Omni NLOS PLE of 1.78 at 6.75 GHz vs. 2.74 at 142 GHz) indicating MPC-rich propagation at FR3.

3GPP does not include a CI FSPL model with a 1 m reference distance option for path loss in InF environments; instead, it uses the Floating Intercept (FI) model with non-physical parameters that can vary over many orders of magnitude, as specified in \textit{Table 7.4.1-1: Path Loss Models} in \cite{3GPPTR38901}. The FI model varies the y-intercept as a fitting parameter and does not use a physical reference of the free-space propagation~\cite{rappaport2017investigation}. Furthermore, 3GPP classifies the factory environment for TX heights above the surrounding clutter as sparse clutter-high height (SH) and dense clutter-high height (DH). Model parameters for both classifications are shown in Table \ref{tab:PLEs}.

\renewcommand{\arraystretch}{1.2}
\begin{table}[!t]
		\centering
		\caption{RMS Delay Spread Characteristics at 6.75 GHz and 16.95 GHz with Comparison to 3GPP~\cite{3GPPTR38901} and 142 GHz InF Measurements in \cite{Ju2023twc}.}
		\label{tab:RMS_DS}
	\resizebox{0.44\textwidth}{!}{	\begin{tabular}{|p{2.2 cm}|p{0.5 cm}|p{1.3 cm}|p{0.5 cm}|p{1.3 cm}|p{0.5 cm}|}
			\hline			\textbf{Frequency (GHz)} & \textbf{6.75} & \textbf{6.75 (3GPP)~\cite{3GPPTR38901}} & \textbf{16.95} & \textbf{16.95 (3GPP)~\cite{3GPPTR38901}} &  \textbf{142\cite{Ju2023twc}} \\ \hline
			\multicolumn{6}{|c|}{Dir RMS DS}     \\ \hline
			LOS \, $\mathbb{E}(\cdot)$ (ns) & 11.2    &--          & 13.7   &--            & 1.8                         \\ \hline
			NLOS \, $\mathbb{E}(\cdot)$ (ns) & 11.3    &--          & 9.2   &--            & 3.2                        \\ \hline
			\multicolumn{6}{|c|}{Omni RMS DS}          \\ \hline
			LOS \, $\mathbb{E}(\cdot)$ (ns) & 14     &26.84          & 12.7   &26.84            & 5.5                        \\ \hline
			NLOS \, $\mathbb{E}(\cdot)$ (ns)& 30.3   &30.92           & 29.0   &30.92            & 11                        \\ \hline
		\end{tabular}
		}\vspace{-1\baselineskip}
        \vspace{-5 pt}
	\end{table}
\renewcommand{\arraystretch}{1}

\setlength{\tabcolsep}{10pt} 
\begin{table*}[!t]
	\centering
	\caption{Angular Spread Characteristics measured by NYU WIRELESS at 6.75 GHz and 16.95 GHz in LOS and NLOS for the InF environment and Comparison to 3GPP Models \cite{3GPPTR38901}}
	\label{tab:AS}
	\renewcommand{\arraystretch}{1.4} 
	\resizebox{0.98\textwidth}{!}{
		\begin{tabular}{|>{\centering\arraybackslash}m{1.9cm}|>{\centering\arraybackslash}m{1.15cm}|>{\centering\arraybackslash}m{1cm}|>{\centering\arraybackslash}m{1.4cm}|>{\centering\arraybackslash}m{1cm}|>{\centering\arraybackslash}m{1cm}|>{\centering\arraybackslash}m{1.15cm}|>{\centering\arraybackslash}m{1.4cm}|>{\centering\arraybackslash}m{1cm}|>{\centering\arraybackslash}m{1cm}|}
			\specialrule{1.2 pt}{0 pt}{0 pt}
		
		\multirow{2}{*}{\textbf{Metric}} & \multirow{2}{*}{\textbf{Condition}} & \multicolumn{4}{c|}{\textbf{6.75 GHz}} & \multicolumn{4}{c|}{\textbf{16.95 GHz}} \\
		\cline{3-10}
		& & \multicolumn{1}{c|}{\textbf{NYU~\cite{Ying2025tcom}}} & \multicolumn{1}{c|}{\textbf{3GPP~\cite{3GPPTR38901}}} & \multicolumn{1}{c|}{$\mathbb{E}(\textbf{NYU})$} & \multicolumn{1}{c|}{$\mathbb{E}(\textbf{3GPP})$} & \multicolumn{1}{c|}{\textbf{NYU~\cite{Ying2025tcom}}} & \multicolumn{1}{c|}{\textbf{3GPP~\cite{3GPPTR38901}}} & \multicolumn{1}{c|}{$\mathbb{E}(\textbf{NYU})$} & \multicolumn{1}{c|}{$\mathbb{E}(\textbf{3GPP})$} \\
		\specialrule{1 pt}{0 pt}{0 pt}
		
		\multirow{4}{*}{\shortstack{Omni RMS $\lg_\text{ASA}$\\  $=\log_{10}$($\text{ASA}/1^{\circ}$)}}
		& $\mu_\text{lgASA}^\text{LOS}$& 1.25 & 1.62 & \multirow{2}{*}{23.71$^\circ$} & \multirow{2}{*}{46.56$^\circ$} & 1.05 & 1.55 & \multirow{2}{*}{14.32$^\circ$} & \multirow{2}{*}{40.86$^\circ$} \\ \cline{2-3} \cline{4-4} \cline{7-8} 
		& $\sigma_\text{lgASA}^\text{LOS}$ & 0.50 & 0.31 &  &  & 0.46 & 0.35 &  &  \\ \cline{2-10}
		& $\mu_\text{lgASA}^\text{NLOS}$ & 1.80 & 1.72 & \multirow{2}{*}{66.38$^\circ$} & \multirow{2}{*}{58.21$^\circ$} & 1.61 & 1.72 & \multirow{2}{*}{50.40$^\circ$} & \multirow{2}{*}{58.21$^\circ$} \\ \cline{2-3} \cline{4-4} \cline{7-8} 
		& $\sigma_\text{lgASA}^\text{NLOS}$ & 0.21 & 0.30 &  &  & 0.43 & 0.30 &  &  \\ \specialrule{1 pt}{0 pt}{0 pt}

		\multirow{4}{*}{\shortstack{Omni RMS $\lg_\text{ASD}$\\=$\log_{10}$($\text{ASD}/1^{\circ}$)}}
		& $\mu_\text{lgASD}^\text{LOS}$ & 1.43 & 1.56 & \multirow{2}{*}{35.89$^\circ$} & \multirow{2}{*}{39.02$^\circ$} & 1.34 & 1.56 & \multirow{2}{*}{22.91$^\circ$} & \multirow{2}{*}{39.02$^\circ$} \\ \cline{2-3} \cline{4-4} \cline{7-8} 
		& $\sigma_\text{lgASD}^\text{LOS}$ & 0.50 & 0.25 &  &  & 0.20 & 0.25 &  &  \\ \cline{2-10}
		& $\mu_\text{lgASD}^{\text{NLOS}}$ & 1.64 & 1.57 & \multirow{2}{*}{45.93$^\circ$} & \multirow{2}{*}{38.90$^\circ$} & 1.49 & 1.57 & \multirow{2}{*}{34.77$^\circ$} & \multirow{2}{*}{38.90$^\circ$} \\ \cline{2-3} \cline{4-4} \cline{7-8} 
		& $\sigma_\text{lgASD}^\text{NLOS}$ & 0.21 & 0.20 &  &  & 0.32 & 0.20 &  &  \\ 
		\specialrule{1.2 pt}{0 pt}{0 pt}
	\end{tabular}}  
	\vspace{-5pt}
\end{table*}
\renewcommand{\arraystretch}{1}

\subsection{RMS Delay Spread}
The RMS DS provides a measure of the temporal spread of MPCs in an environment. A power threshold of 25 dB below the peak or 5 dB above the noise floor is used to evaluate the RMS DS. Table \ref{tab:RMS_DS} presents the RMS DS at 6.75 GHz and 16.95 GHz, along with the 3GPP results from \textit{Table 7.5-6 Part-3: Channel Model Parameters for InF} in \cite{3GPPTR38901}. 
For a direct comparison with the log-normally distributed 3GPP RMS DS models, the RMS DS for the InF environment is evaluated by first taking the $\log_{10}$ of the RMS DS from the omni PDP at each TX-RX location~\cite{3GPPTR38901}. The $\log_{10}$ operation converts the Omni RMS DS data into a normally distributed data, allowing the computation of arithmetic mean ($\mu$) and sample standard deviation ($\sigma$) of the log-transformed RMS DS values and a direct comparison with 3GPP models \cite{3GPPTR38901}. In simpler words, one would calculate the arithmetic mean by taking the values in the ``Omni DS'' column of Table \ref{tab:LSPs} for a measurement frequency and calculating $\mu=\frac{\sum_{i=1}^{N}\text{ }\log_{10}(\text{Omni DS}_i)}{N}$, where OmniDS$_i$ is the Omni RMS DS at each TX-RX location and $N$ is the total number of measurement locations. Similarly, sample standard deviation is compute as $\sigma = \sqrt{\frac{\sum_{i=1}^{N}(\log_{10}(\text{Omni DS}_i)-\mu)^2}{N-1}}$, which provides an unbiased estimate of the standard deviation. The $\mu$ and $\sigma$ are obtained for the normally distributed data and to express the distribution back into a linear nanosecond scale, the expectation is calculated using $\mu$ and $\sigma$  ($\mathbb{E}(\text{RMS DS}) = 10^{\mu+\frac{\sigma^2}{2}}$).

The RMS DS in the 3GPP model depends on the physical dimensions of the environment, particularly the volume (V) and enclosing surface area (S) considering the walls, ceiling, and floor, rather than the frequency \cite{3GPPTR38901}. For the NYU MakerSpace floorplan shown in Fig. \ref{fig:IndoorMap}, the V is calculated as 3470 $\text{m}^3$ and the S as 1958 $\text{m}^2$, assuming a uniform ceiling height of 5 m.

\renewcommand{\arraystretch}{1.1}
\begin{table*}[!t]
	\centering
 \color{black}
		\vspace{10pt}
        \caption{Point-data table for site-specific (InF) large scale spatio-temporal statistics with map in Fig. \ref{fig:IndoorMap} \cite{Ted2025icc,Ying2025tcom} }
        \resizebox{1\textwidth}{!}{
	\begin{tabular}{@{\hspace{10 pt}}p{0.35 cm}p{0.3 cm}@{\hspace{10 pt}}p{0.4 cm}@{\hspace{10 pt}}p{0.4 cm}p{0.4 cm}p{0.7 cm}@{\hspace{10 pt}}p{0.7 cm}p{0.4 cm}p{0.4 cm}p{0.4 cm}p{0.4 cm}p{0.4 cm}p{0.4 cm}p{0.4 cm}p{0.4 cm}p{0.4 cm}p{0.4 cm}}
		\hline
		\multicolumn{1}{p{0.35 cm}}{\textbf{Freq.}} & \textbf{TX} & \textbf{RX} & \textbf{Loc.} & \multicolumn{1}{p{0.4 cm}}{\textbf{TR Sep.}} &\textbf{Omni Abs. PL (V-V)} & \textbf{Omni Abs. PL (V-H)} & \multicolumn{1}{p{0.4 cm}}{\textbf{Mean Dir. DS}} & \multicolumn{1}{p{0.4 cm}}{\textbf{Omni DS}} & \multicolumn{1}{p{0.4 cm}}{\textbf{Mean Lobe ASA}} & \multicolumn{1}{p{0.4 cm}}{\textbf{Omni ASA}} & \multicolumn{1}{p{0.4 cm}}{\textbf{Mean Lobe ASD}} & \multicolumn{1}{p{0.4 cm}}{\textbf{Omni ASD}} & \multicolumn{1}{p{0.4 cm}}{\textbf{Mean Lobe ZSA}} & \multicolumn{1}{p{0.4 cm}}{\textbf{Omni ZSA}} & \multicolumn{1}{p{0.4 cm}}{\textbf{Mean Lobe ZSD}} & \multicolumn{1}{p{0.4 cm}}{\textbf{Omni ZSD}} \\
		\hline
		\text{[GHz]}& & & &[m]&[dB]&[dB]&[ns]&[ns]&[$^\circ$]&[$^\circ$]&[$^\circ$]&[$^\circ$]&[$^\circ$]&[$^\circ$]&[$^\circ$]&[$^\circ$] \\ 
		\hline
			\multirow{12}{*}{6.75} & \multirow{5}{*}{TX1} & RX1   & LOS   & 14.6  & 67.04 & 80.8  & 5.9   & 26.7  & 6.1   & 55.2  & 11.4  & 31.3  & 13.5  & 18.2  & 11.1  & 10.7 \\
          &       & RX2   & NLOS  & 16.6  & 69.81 & 87.44 & 8.8   & 32.7  & 7.4   & 50.4  & 11.0  & 26.5  & 7.6   & 14.8  & 4.5   & 11.3 \\
          &       & RX3   & NLOS  & 11.1  & 67.6  & 84.32 & 7.0   & 23.0  & 83.8  & 83.8  & 14.5  & 24.4  & 24.5  & 24.5  & 2.6   & 9.8 \\
          &       & RX4   & LOS   & 9.0   & 63.12 & 77.38 & 13.2  & 8.0   & 5.8   & 5.8   & 5.9   & 12.3  & 5.8   & 17.9  & 5.9   & 13.4 \\
          &       & RX5   & NLOS  & 19.0  & 71.38 & 83.26 & 5.6   & 27.3  & 124.0 & 124.0 & 18.3  & 28.2  & 1.1   & 7.6   & 2.9   & 9.2 \\
          \cline{2-17}
          & \multirow{4}{*}{TX2} & RX1   & LOS   & 33.3  & 72.02 & 89.56 & 3.0   & 14.6  & 7.9   & 62.1  & 13.7  & 41.6  & 9.3   & 20.4  & 1.8   & 8.9 \\
          &       & RX2   & NLOS  & 38.2  & 74.82 & 96.41 & 4.3   & 27.9  & 19.4  & 75.8  & 10.4  & 74.7  & 3.9   & 13.8  & 4.1   & 11.9 \\
          &       & RX3   & LOS   & 24.9  & 65.55 & 87.68 & 7.5   & 6.1   & 6.1   & 6.1   & 5.9   & 6.8   & 6.1   & 19.0  & 5.9   & 13.4 \\
          &       & RX4   & NLOS  & 24.6  & 79.06 & 96.65 & 13.5  & 40.5  & 47.7  & 47.7  & 48.2  & 48.2  & 12.6  & 12.6  & 1.7   & 5.7 \\
          \cline{2-17}
          & \multirow{3}{*}{TX3} & RX1   & LOS   & 19.3  & 65.52 & 84.35 & 10.2  & 18.7  & 15.2  & 15.2  & 15.9  & 133.3 & 2.4   & 13.4  & 6.1   & 11.3 \\
          &       & RX2   & NLOS  & 9.5   & 63.62 & 82.58 & 7.6   & 24.6  & 28.5  & 28.5  & 47.0  & 57.3  & 9.5   & 13.0  & 2.7   & 8.8 \\
          &       & RX3   & NLOS  & 14.0  & 69.85 & 87.58 & 4.8   & 38.6  & 26.6  & 78.4  & 72.0  & 75.9  & 13.1  & 13.3  & 2.3   & 7.0 \\
          \hline
          \hline
    \multirow{12}{*}{16.95} & \multirow{5}{*}{TX1} & RX1   & LOS   & 14.6  & 74.07 & 99.07 & 13.2  & 17.3  & 7.7   & 7.7   & 7.2   & 17.2  & 2.1   & 7.6   & 4.1   & 7.0 \\
          &       & RX2   & NLOS  & 16.6  & 78.09 & 104   & 11.3  & 56.1  & 7.5   & 30.7  & 9.5   & 18.6  & 5.4   & 9.6   & 3.9   & 6.4 \\
          &       & RX3   & NLOS  & 11.1  & 79.33 & 99.46 & 6.0   & 41.2  & 9.3   & 69.4  & 9.9   & 17.0  & 10.8  & 11.1  & 2.8   & 6.1 \\
          &       & RX4   & LOS   & 9.0   & 73.08 & 99.45 & 10.4  & 8.6   & 6.7   & 6.7   & 7.0   & 11.7  & 9.3   & 9.3   & 4.0   & 6.8 \\
          &       & RX5   & NLOS  & 19.0  & 88.54 & 116.5 & 3.3   & 2.8   & 5.8   & 5.8   & 8.8   & 10.5  & 8.9   & 11.5  & 2.8   & 5.9 \\
          \cline{2-17}
          & \multirow{4}{*}{TX2} & RX1   & LOS   & 33.3  & 82.63 & 105.7 & 8.5   & 22.0  & 6.8   & 71.6  & 9.6   & 33.6  & 7.1   & 11.8  & 2.0   & 5.5 \\
          &       & RX2   & NLOS  & 38.2  & 86.96 & 111.9 & 1.1   & 11.3  & 7.0   & 75.9  & 7.3   & 60.7  & 5.7   & 13.1  & 5.4   & 9.9 \\
          &       & RX3   & LOS   & 24.9  & 87.31 & 112.8 & 5.5   & 4.7   & 5.8   & 5.8   & 5.6   & 20.9  & 5.4   & 11.5  & 5.2   & 8.5 \\
          &       & RX4   & NLOS  & 24.6  & 90.65 & 100.7 & 9.0   & 43.8  & 19.5  & 64.7  & 14.5  & 65.6  & 2.8   & 6.0   & 3.6   & 5.3 \\
          \cline{2-17}
          & \multirow{3}{*}{TX3} & RX1   & LOS   & 19.3  & 77.97 & 98.77 & 10.9  & 13.9  & 8.0   & 8.0   & 6.9   & 35.3  & 2.2   & 7.8   & 1.6   & 5.1 \\
          &       & RX2   & NLOS  & 9.5   & 76.09 & 102.9 & 5.7   & 17.4  & 6.1   & 29.7  & 7.6   & 34.9  & 4.6   & 10.8  & 2.8   & 5.9 \\
          &       & RX3   & NLOS  & 14.0  & 82.15 & 105.6 & 3.9   & 50.5  & 14.1  & 106.0 & 15.3  & 58.5  & 5.9   & 7.7   & 3.8   & 6.9 \\
			\hline
		\end{tabular}} 
		\label{tab:LSPs}%
		\vspace{-2 pt}
	\end{table*}%
	\medskip
	\renewcommand{\arraystretch}{1}

\begin{figure}[!t]
	\centering%
	\subfloat[]{
		\centering
		\includegraphics[width=45mm]{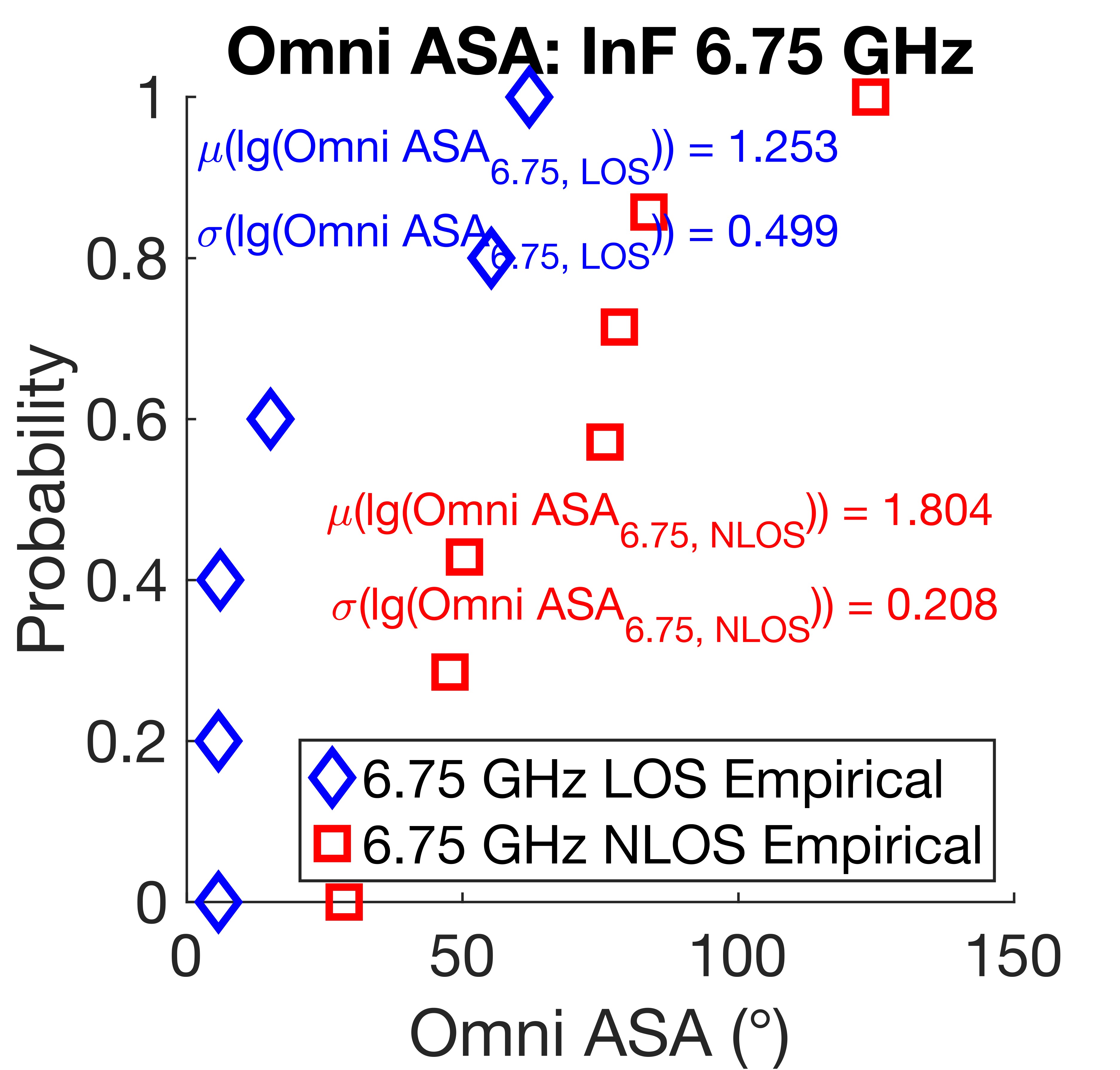}
	}
	\subfloat[]{
		\centering
		\includegraphics[width=45mm]{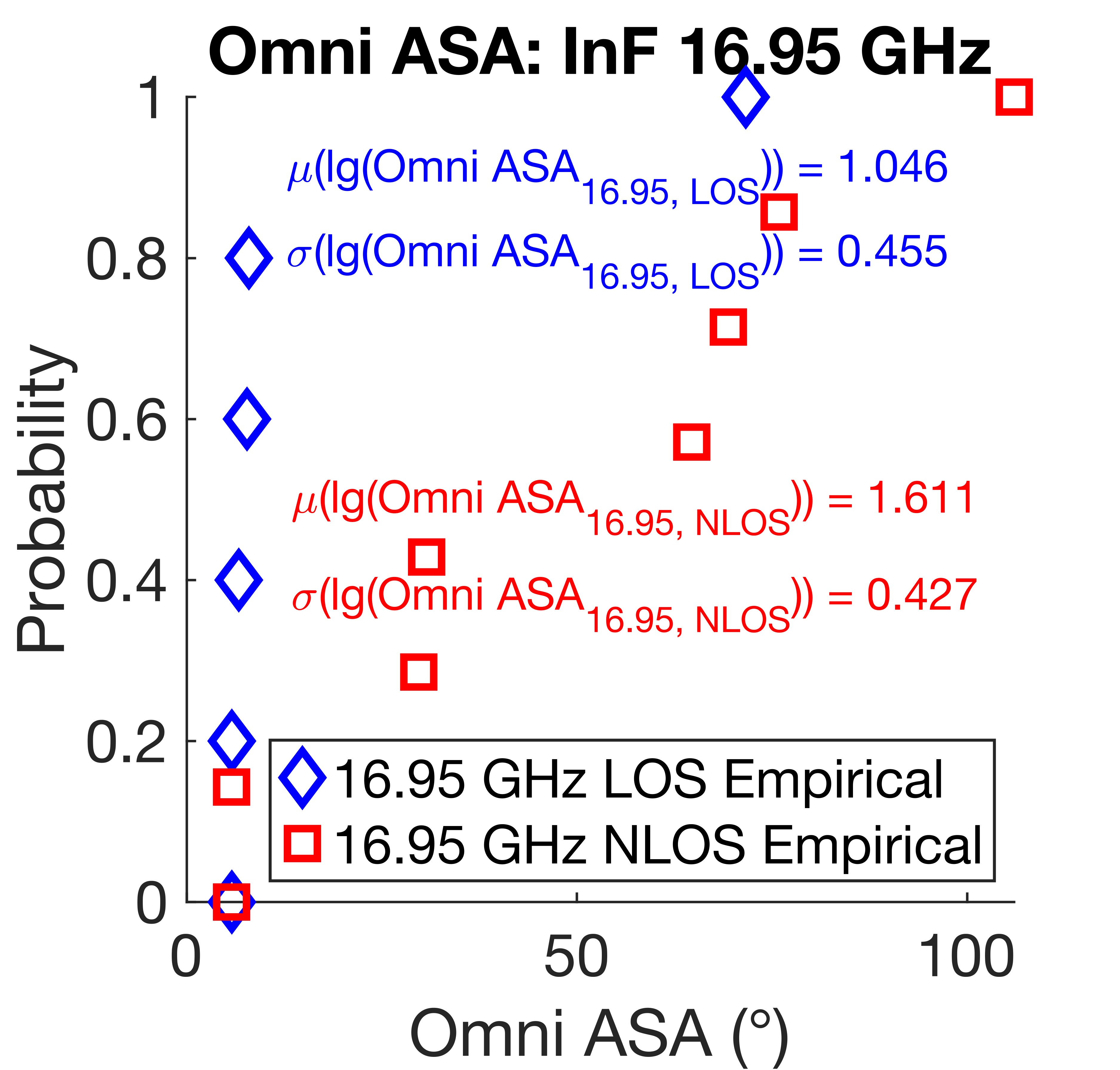}
	}%
	\\
	\caption{InF omni ASA: (a) 6.75 GHz; (b) 16.95 GHz. [T-R separation: 9--38 m]}
	\label{fig:ASA_D}
	\vspace{-10 pt}
\end{figure}

\subsection{RMS Angular Spread}

The RMS AS shows the spatial dispersion of arrival or departure angles of MPCs in azimuth and elevation planes. RMS AS is evaluated using a spatial lobe threshold of 10 dB below the strongest pointing direction, using the 3GPP method detailed in \cite{3GPPTR38901}. The AS is commonly evaluated using the 3GPP method \cite{3GPPTR38901} (circular standard deviation of MPCs) or Fleury's method (second central moment of the PAS) \cite{Ying2025tcom}. A comprehensive comparison between the 3GPP model and conventional Fleury's method for computing AS is presented in \cite{Ying2025tcom}. Similarly, by preserving the underlying log-normal distribution of the measured AS, we compare our AS measurements with the 3GPP model using the same approach as that applied to DS, as explained in Section III. B. Table \ref{tab:AS} summarizes the AS observed for arrival angles at the RX (ASA) and departure angles at the TX (ASD). In LOS, the ASA is observed as 23.71$^\circ$ at 6.75 GHz and 14.32$^\circ$ at 16.95 GHz. Wider ASAs of 66.38$^\circ$ and 50.40$^\circ$ are observed in NLOS at 6.75 GHz and 16.95 GHz, respectively. 
3GPP uses a frequency-dependent LOS ASA with overestimated spread, along with a constant NLOS ASA, as outlined in \textit{Table 7.5-6 Part-3: Channel Model Parameters for InF} and \textit{Table 7.5-11: ZSD and ZOD Offset Parameters for InF} in \cite{3GPPTR38901}. Measured ASAs are observed to decrease with frequency in both LOS and NLOS at 6.75 GHz and 16.95 GHz.
Further, we found ASDs to be 35.89$^\circ$ in LOS and 45.93$^\circ$ in NLOS at 6.75 GHz, while ASDs of 22.91$^\circ$ in LOS and 34.77$^\circ$ in NLOS were observed at 16.95 GHz. As shown in Fig. \ref{fig:ASA_D}, the CDFs of empirical data align with these trends, while 3GPP predictions suggest a constant ASD of 39$^\circ$ in LOS and 38.9$^\circ$ in NLOS from 0.5 GHz to 100 GHz. 


Point-data for large-scale spatio-temporal parameters corresponding to each TX-RX location pair are summarized in Table \ref{tab:LSPs} in concert with Fig. \ref{fig:IndoorMap}. This point-data format enables pooling measurement data from multiple contributors by simply appending rows to the table. 
Such point-data table can be used to derive traditional statistical models by computing CDFs and scatter plots down each column, such as Fig. \ref{fig:ASA_D}, while simultaneously preserving the site-specific nature of each measurement location for developing ray-tracing models or applying AI/ML techniques without any additional information other than that denoted in the columns of Table \ref{tab:LSPs} \cite{Ted2025icc,Ying2025tcom}.

\section{Conclusion}
\label{conclusion}
We presented an in-depth analysis of radio propagation behavior for InF at 6.75 GHz and 16.95 GHz, based on extensive measurements. 12 TX-RX locations were measured, encompassing five LOS and seven NLOS locations spanning 9- 38 m. The path loss analysis revealed significant received power even in NLOS scenarios at 6.75 GHz. The PLEs were observed to be small with values below 2 in LOS and close to 2 in NLOS at both 6.75 and 16.95 GHz. The RMS DS was observed to decrease with increasing frequencies in both LOS and NLOS. The captured RMS AS also showed a decreasing trend with frequency, and wide angular spreads were observed particularly in NLOS. Frequency-dependent trends for DS, and AS were observed in contrast to some of the constant modeling currently provided in 3GPP specifications.

\bibliographystyle{IEEEtran}
\bibliography{references}

\end{document}